\documentclass[a4paper]{jpconf}
\usepackage{graphicx}
\usepackage{subcaption}
\usepackage{comment}
\usepackage{amsmath}
\usepackage{amsfonts}
\usepackage{cleveref} 
\usepackage{svg}
\usepackage{siunitx}
\usepackage{pdfpages}
\usepackage{tabularx}

\newcolumntype{L}{>{\raggedright\arraybackslash}p{0.36\textwidth}}
\newcolumntype{M}{>{\raggedright\arraybackslash}p{0.3\textwidth}}
\newcolumntype{N}{>{\raggedright\arraybackslash}p{0.24\textwidth}}

\begin{document}
\title{Probabilistic surrogate modeling of offshore wind-turbine loads with chained Gaussian processes} 
\author{D Singh$^1$, R P Dwight$^1$, K Laugesen$^2$, L Beaudet$^3$, 
and A Vir\'e$^1$}
\address{$^1$ Aerodynamics, Wind Energy and Propulsion, TU Delft, Kluyverweg 1, 2629 HS Delft, The Netherlands}
\address{$^2$ Siemens Gamesa Renewable Energy, Borupvej 16, 7330 Brande, Denmark}
\address{$^3$ Siemens Gamesa Renewable Energy, Avenue de l'Universit\'e, 76800 Saint-Etienne-du-Rouvray, France}
\ead{d.singh-1@tudelft.nl}


\begin{abstract}
Heteroscedastic Gaussian process regression, based on the concept of chained Gaussian processes, is used to build surrogates to predict site-specific loads on an offshore wind turbine. Stochasticity in the inflow turbulence and irregular waves results in load responses that are best represented as random variables rather than deterministic values. Moreover, the effect of these stochastic sources on the loads depends strongly on the mean environmental conditions -- for instance, at low mean wind speeds, inflow turbulence produces much less variability in loads than at high wind speeds. Statistically, this is known as {\it heteroscedasticity}. Deterministic and most stochastic surrogates do not account for the heteroscedastic noise, giving an incomplete and potentially misleading picture of the structural response. In this paper, we draw on the recent advancements in statistical inference to train a heteroscedastic surrogate model on a noisy database to predict the conditional pdf of the response. The model is informed via $10$-minute load statistics of the IEA-\SI{10}{\mega\watt}-RWT subject to both aero- and hydrodynamic loads, simulated with OpenFAST. Its performance is assessed against the standard Gaussian process regression. The predicted mean is similar in both models, but the heteroscedastic surrogate approximates the large-scale variance of the responses significantly better.

\end{abstract}

\section{Introduction}
\label{sec:introduction}
Prior to installing a wind turbine at a new site, one must thoroughly analyze its structural integrity over its lifetime. The site assessment phase typically involves many mid-to high-fidelity coupled aero-hydro-servo- elastic simulations with site-specific environmental inputs to calculate the fatigue and extreme loads on the wind turbine. However, the complex numerical models for evaluating offshore wind turbines can often be computationally expensive, and the simulations in the design load basis are prohibitively numerous. For instance, simulations made at every \SI{2}{\meter\per\second} as dictated by the certification guidelines \cite{IEC_fixed} lead to 10 simulations in the wind speed dimension. With $n$ such parameters significantly impacting fatigue loads, the total number of simulations can quickly scale up to $10^n$ \cite{Muller2018ApplicationTurbines}. In such cases, we may use low-cost statistical models called \emph{surrogate models} that map the mean environmental conditions (here denoted $\boldsymbol x$) to the loads  ($y$) to reduce the computational overhead. 

In nature, offshore wind turbines are subject to randomly varying turbulence and waves, whose statistical properties depend on $\boldsymbol x$. This source of stochasticity is mimicked in deterministic numerical simulators via pseudo-random number generators, initialized by random seeds. That is, with the same $\boldsymbol x$ but different random seeds, the simulation code produces unique turbulence and wave time-series, and thereby, unique load time-series. Whether the training is done on measured- or numerical data, the noise due to these random events is an irreducible quantity, i.e., \ an \emph{aleatoric} uncertainty. Furthermore, the variance of this noise is a function of the wind speed, wave period, and turbulence intensity, among other environmental conditions. The statistical term for such input-dependent noise is \emph{heteroscedastic} noise. As such, deterministic surrogates that fit functions of the form $\boldsymbol{x} \mapsto y$ give an incomplete picture of the simulation response. Additionally, fitting a flexible, highly non-linear model on a large training dataset can lead to the model trying to fit the noise instead of the latent mean of the response. On the other hand, a too-small training dataset may result in a model that does not meaningfully learn to predict point estimates.

One way of dealing with the stochasticity in the loads is to average them over a very long time period or average over several repetitions of the $10$-minute simulations with different random seeds before training the surrogate. The IEC61400-1 standard recommends a $60$-minute long time signal or six $10$-minute simulations. However, $60$ minutes of data has also been shown to be insufficient to characterize the uncertainty in the fatigue loads fully \cite{Muller2018ApplicationTurbines, Zwick2015TheDaniel}.

Alternative approaches model the distribution (e.g.,\ via the probability density function (pdf)) from which the training samples are drawn directly. Broadly such methods can be based on either (i) replication, in which multiple random seeds are evaluated at each $\boldsymbol x$, or on (ii) models which rely on the smoothness of the response in order to estimate statistics which vary in $\boldsymbol x$. The latter - as the methods presented in this paper - approximate the conditional distribution of the response, given the data, i.e., $y\mid\boldsymbol x$. They can infer the underlying hyper-parameters of the pdf directly from a noisy data set within the bounds of assumptions specified by the user. 
 
 
 

Wind turbine loads have been modeled using a vast range of data-driven methods, of which probabilistic models form a relatively small part. Zhu et al. \cite{Zhu2020Replication-basedDistributionsb} introduce joint polynomial chaos expansion-generalized $\lambda$-distribution (PCE-GLD) algorithm to model the pdf of the response. The parameters of the $\lambda$-distribution are calibrated from the data using the maximum likelihood estimate. The model accurately predicts the pdf for a simple test case of a fixed-bottom wind turbine with aerodynamic loading. The main advantage of this method is that it does not assume a Gaussian distribution for the responses and accounts for the heteroscedasticity in the noise. Abdallah et al. \cite{Abdallah2019ParametricTurbines} use parametric hierarchical Kriging to predict blade-root-bending-moment extreme loads that are heteroscedastic on a \SI{2}{\mega\watt} onshore wind turbine. Their approach combines low- and high-fidelity observations, where the low-fidelity model informs the high-fidelity Gaussian Process regression (GPR). They show that introducing hierarchy helps make the model selection process more robust than the manual tuning of Kriging parameters. Fatigue loads on the tower in the presence of wind and wave loads have been previously modeled using standard GPR by Texeira et al. \cite{Teixeira2017AnalysisSurfaces}; however, they omit heteroscedasticity. GPR has been compared to other data-driven methods like linear regression, and artificial neural networks by Gasparis et al. \cite{Gasparis2020SurrogateFarm} for modeling power and fatigue loads, showing a superior performance by the GPR. Similarly, Dimitrov et al. \cite{Dimitrov2018FromDatabases} evaluate importance sampling, nearest-neighbor interpolation, polynomial chaos expansion (PCE), GPR, and quadratic response surface (QRS), to conclude a better performance again by the GPR despite a computational penalty.

Evidently, wind turbine load emulation is an active area of research, and several data-driven approaches have been explored to tackle this issue. However, very few methods focus on uncertainty quantification and consider the heteroscedastic nature of the problem. Statistical inference, specifically, has not been used to model high-dimensional, multi-fidelity load surrogates for offshore wind turbines with aero-hydrodynamic loading. This paper aims to assess the performance of the heteroscedastic  Gaussian process regression (H-GPR) based on the recently introduced chained Gaussian processes (CGPs) \cite{Saul2016ChainedProcesses}. We test the model's performance in a six-dimensional space of environmental parameters ($\boldsymbol x$), with aleatoric uncertainty due to random turbulence and wave sources, by comparing the conditional distribution of the surrogate with repetitive sampling from the full-order model and with standard Gaussian process regression. Overall, H-GPR shows a considerable improvement over GPR in predicting the conditional distribution of the average and maximum loads. The standard deviation of loads is more challenging to predict, especially at low wind speeds at the tower base, where it is driven by hydrodynamic conditions -- nonetheless, H-GPR performs well considering the limited data.

The paper is structured in the following manner; \cref{sec:methodology} briefly describes the probabilistic modeling approach with \cref{subsec:methodology_gpr} and \cref{subsec:methodology_cgpr} talking about GPR and H-GPR, respectively. The setup of the full order model along with the machine learning framework is described in \cref{sec:setup}. The results assessing the performance of the two models are presented in \cref{sec:results}.
\section{Methodology}
\label{sec:methodology}

Probabilistic approaches directly account for the irreducible uncertainty in the system by modeling either the latent variables causing noise or the noise itself. The latent quantities are inferred based on the observed data using principles of probabilistic inference. Suppose we have a noisy data set with vectors of random variables to denote the inputs $\boldsymbol X : \Omega \rightarrow \mathbb{R}^{M}$ (the mean environmental conditions), response $Y : \Omega \rightarrow \mathbb{R}$ (a single mean wind-turbine load channel) and the aleatoric uncertainty sources $\boldsymbol \varTheta:\Omega \rightarrow \mathbb{R}^{P}$. Throughout $\Omega$ is a suitable probability space. The observations $y_i\in\mathbb{R}$ at $\boldsymbol x_i\in\mathbb{R}^M$, $\boldsymbol{\theta}_i\in\mathbb{R}^P$ are samples of the joint random variable $(\boldsymbol{X}, \boldsymbol{\varTheta}, Y)$ where we assume $\boldsymbol{X} \perp \boldsymbol{\varTheta}$. We are interested in a surrogate model of the form $Y\mid \boldsymbol{x} \simeq f(\boldsymbol x; \boldsymbol \varTheta)$ for some function $f$.

Typical deterministic surrogates $\hat y$, aim to approximate the mean of $Y\mid \boldsymbol{x}$:
\begin{equation*}
        \hat y(\boldsymbol x) \simeq \mathbb{E}_{\boldsymbol \varTheta} (\boldsymbol Y \mid \boldsymbol x),
\end{equation*}
averaging the inherent variability of the loads at a given vector of mean environmental conditions, thereby potentially misrepresenting the frequency and intensity of extreme loads. In contrast, probabilistic models $\hat Y(\boldsymbol x)$ aim to infer the  conditional distribution directly:
\begin{equation*}
    \hat Y(\boldsymbol x) \approx Y \mid \boldsymbol x,
\end{equation*}
encompassing the statistical dependence on $\boldsymbol \varTheta$ and providing a complete picture of expected loads. Bayesian/probabilistic inference provides an elegant solution for identifying $\hat Y(\boldsymbol x)$ from data. Probabilistic inference works on a basic principle of placing hypotheses on the model parameters (\emph{priors}) and calibrating the hypotheses based on how close the model predictions are to the observations (\emph{likelihood}). As a first step, we must choose a model form.

\subsection{Gaussian Process Regression}
\label{subsec:methodology_gpr}

Gaussian process (GP) models are non-parametric, flexible, Bayesian machine learning models widely used for regression problems. A brief overview of Gaussian process regression, based on  \cite{books/lib/RasmussenW06} is given in this section. 
Assuming a system with an $M$ dimensional input vector $\boldsymbol{x} \in \mathbb{R}^M$ and an additive noise element $\varepsilon$, the regression model can be defined as:
\begin{align}
\label{eq:gpr_model}
    y = f(\boldsymbol{x}) + \varepsilon, \quad
    \varepsilon \sim \mathcal N(0, \sigma^2).
\end{align}
The noise is modeled as an independent Gaussian distribution centered around $0$ with a variance of $\sigma^2$ which we may also estimate from the data. We are interested in evaluating the predictive distribution $p(y^\star \mid \textbf{y}, X, \boldsymbol{x^{\star}})$, that estimates the function value $ y^{\star}$ given a unobserved parameter vector $ \boldsymbol{x^{\star}}$ and $N$ observations $X \in \mathbb{R}^{N\times M}$ and $\boldsymbol{y} \in \mathbb{R}^{N}$. A Gaussian process prior is assumed for the latent function $f(\boldsymbol x) \sim \mathcal{GP}(\mu(\boldsymbol x), k(\boldsymbol{x}, \boldsymbol{x'}))$.
The covariance kernel $k$ dictates the smoothness of the function -- in this paper, we use the squared exponential kernel 
\begin{equation*}
    k(\boldsymbol{x, x'}) = \sigma^2_h \exp \left( -\frac{1}{2} \| \boldsymbol{x-x'} \|^{2}_{\boldsymbol{l}} \right), \quad \| \boldsymbol{x-x'} \|^{2}_{\boldsymbol{l}} = \sum_{j=1}^M \frac{|x^{(j)} - x'^{(j)}|^2}{l^{(j)}},
\end{equation*}
implying that the underlying function is smooth and infinitely differentiable, and where $x^{(j)}$ is the $j$-th component of $\boldsymbol{x}$.  The characteristic length-scales $\boldsymbol{l}\in\mathbb{R}^M$ are defined per input parameter, and these and the variance $\sigma^2_h$ are \emph{hyperparameters} that are tuned using the data. 

The observations $\boldsymbol{y}$ and prediction $y^{\star}$ are jointly Gaussian
\begin{gather*}
 \begin{bmatrix} \boldsymbol{y} \\ y^\star \end{bmatrix}
 \sim 
 \mathcal{N} \left( \begin{bmatrix} 
 \mu(X)\\
 \mu(\boldsymbol x^{\star})
 \end{bmatrix},
  \begin{bmatrix}
   K_{XX} + \sigma^2 I &
   K_{X\boldsymbol{x^\star}} \\
   K_{\boldsymbol{x^\star} X} &
    K_{\boldsymbol{x^\star x^\star}} + \sigma^2 I 
   \end{bmatrix}
   \right).
\end{gather*}
The joint distribution is conditioned on the observed values to get the predictive distribution corresponding to a new input $\boldsymbol{x^\star}$ as,
\begin{align}
    y^\star &\mid \boldsymbol{y}, X, \boldsymbol{x^{\star}} \sim \mathcal{N}(\hat\mu(x^\star), \hat \Sigma(x^\star)) \\
    \hat\mu(x^\star) &= \mu(\boldsymbol x^{\star}) K_{\boldsymbol{x^\star} X} (K_{XX} + \sigma^2 I) ^{-1} (\boldsymbol{y}-\mu(X)) \\
    \hat\Sigma(x^\star) &= K_{\boldsymbol{x^\star x^\star}} - K_{\boldsymbol{x^\star} X} (K_{XX} + \sigma^2 I)^{-1} K_{X\boldsymbol{x^\star}} +\sigma^2
\end{align}
The only missing information to make predictions is the value of the hyperparameters $\sigma_h$, $\boldsymbol{l}$. The hyperparameters may be user-defined, but common practice is to infer them from the data using type-II maximum likelihood \cite{books/lib/RasmussenW06}, wherein the negative log marginal likelihood is minimized with respect to the hyperparameters. The negative log marginal likelihood is defined as,
\begin{equation}
     -{\log p(\boldsymbol y| X, \sigma_h, l_h)} = 
     \frac{1}{2} (\boldsymbol y -\mu(X))^{\top} (K_{XX} +\sigma^2 I )^{-1} (\boldsymbol y - \mu(X))
    + \frac{1}{2} \log |K_{XX} +\sigma^2 I | 
    + \frac{n}{2} \log 2\pi
\end{equation}
The optimization is performed using the  L-BFGS-B algorithm.

\subsection{Chained Gaussian Processes}
\label{subsec:methodology_cgpr}
We have seen that in the standard GPR model, the mean of the likelihood is an $\boldsymbol{x}$-dependent latent function while the additive noise element remains constant. Its scope of applicability is therefore limited to homoscedastic processes with noise independent of $\boldsymbol{x}$. A logical generalization of \eqref{eq:gpr_model} is to write $\varepsilon$ as a function of the input variables \cite{Saul2016ChainedProcesses},
\begin{align}
\label{eqn:heteroscedastic_model}
    y = f(\boldsymbol{x}) + \varepsilon(\boldsymbol{x}), \\
        \varepsilon \sim \mathcal{N}(0, \sigma^2 (\boldsymbol{x})),
\end{align}
and to introduce a new latent function $g(\boldsymbol{x})$ to describe the variance of the noise.  Specifically, to guarantee positive values, the variance is assigned a log-GP prior,
\begin{equation}
    y \sim \mathcal {GP} (f(\boldsymbol{x}) , e^{g(\boldsymbol x)})
\end{equation}
where, $f(\boldsymbol{x}) \sim \mathcal{GP}(\mu_f, k_f(\boldsymbol{x,x'}))$ and $g(\boldsymbol{x}) \sim \mathcal{GP}(\mu_g, k_g(\boldsymbol{x,x'}))$ are independent latent processes with individual means and covariances. Contrary to the standard GPR, the posterior is no longer analytically tractable, and must be approximated. One of the emerging approximation methods in the field of statistical inference is \emph{variational inference} (VI) \cite{Blei2017VariationalStatisticians}. In this paper, we use Chained Gaussian processes (CGPs) \cite{Saul2016ChainedProcesses} that use VI to solve \eqref{eqn:heteroscedastic_model}. CGPs build on the idea of generalized linear models (GLMs), which use link functions to relate the mean function of a GP to the mean of any other distribution from the exponential family of distributions. A major drawback of GLMs is that the link function must be invertible and render the latent functions additive Gaussian. CGPs propose a generalized framework that allows for non-linear combination of any number of latent processes. 

In Gaussian processes with $N$ data points, the cost of inference is $\mathcal O(N^3)$, which severely limits its scalability. The CGP implementation that we use also includes a sparse scheme that reduces the computational cost to $\mathcal O(NI^2)$ for $I$ inducing points. The inducing points are used for parametrizing the covariance matrix, and their locations are inferred variationally. The posterior distribution to be inferred is defined as,
\begin{equation}
\label{eq:latent_variables_posterior}
   p(\boldsymbol{f, g, u}_f, \boldsymbol{u}_g | \boldsymbol y) =\frac{p(\boldsymbol {y| f, g, u}_f, \boldsymbol{u}_g ) p(\boldsymbol{f, g, u}_f, \boldsymbol{u}_g)}
    {p(\boldsymbol y)},
\end{equation}
where the functions $\boldsymbol f =f(\boldsymbol x)$ and $\boldsymbol g = g(\boldsymbol x)$ are valued $\boldsymbol{u}_f$ and $\boldsymbol{u}_g$ at inducing point location $\boldsymbol z$. The issue is that the marginal distribution $p(\boldsymbol y)$ is intractable, therefore the exact posterior distribution cannot be determined. It is approximated by a variational distribution $q \in \mathcal Q$ such that,
\begin{equation*}
    p(\boldsymbol{f, g, u}_f, \boldsymbol{u}_g |\boldsymbol y) \approx p(\boldsymbol{f|u}_f)p(\boldsymbol{f|u}_f)q(\boldsymbol{u}_f)q(\boldsymbol{u}_g).
\end{equation*}
The optimization problem reduces to finding the distributions $q(\boldsymbol{u}_f)$ and $q(\boldsymbol {u}_g)$ so that the KL divergence between the approximate and the true distribution can be minimized. The KL divergence cannot be minimized directly, but we can minimize a function that is equal to it up to a constant using a variational bound for the log marginal likelihood,
\begin{equation}
    \log p(\boldsymbol y) \geq \int q(\boldsymbol{f})q(\boldsymbol{g}) \log p(\boldsymbol{y|f, g}) -
    {KL}(q(\boldsymbol{u}_f)||p(\boldsymbol{u}_f)) - 
    {KL}(q(\boldsymbol{u}_g)||p(\boldsymbol{u}_g)),
    \label{eq:elbo}
\end{equation}
where $q(\boldsymbol{f})= \int p(\boldsymbol {f|u}_f) q(\boldsymbol {u}_f) \mathrm{d}\boldsymbol{u}_f$ and $q(\boldsymbol{g})= \int p(\boldsymbol {g|u}_g) q(\boldsymbol {u}_g) \mathrm{d}\boldsymbol{u}_g$.
Saul et al. \cite{Saul2016ChainedProcesses} parametrize $q(\boldsymbol{u}_f)$ and $q(\boldsymbol{u}_g)$ as Gaussian distributions with variational parameters $\boldsymbol{\mu}_f, \boldsymbol{S}_f$ and $\boldsymbol{\mu}_g, \boldsymbol{S}_g$, respectively. A Gaussian prior is placed over $\boldsymbol f$ and $\boldsymbol g$ allowing $q(\boldsymbol{f})$ and $q(\boldsymbol{g})$ to be computed analytically as a function of the variational parameters, which are learnt through optimization. The integral in \eqref{eq:elbo} factorizes over the data points and solved analytically for a Gaussian likelihood.


For both GPR and H-GPR models, the GPFlow package \cite{GPflow2017} available in Python is used. GPFlow is based on TensorFlow, providing variational approximation as the primary tool for statistical inference.

\section{Setup}
\label{sec:setup}

\subsection{Full Order Model}
The surrogate is modeled on the responses of the aero-hydro-servo-elastic code, OpenFAST \cite{openfast_web, openfast_jonkman}. It is a state-of-the-art, multi-physics numerical tool for modeling wind turbines. It can model environmental conditions like stochastic waves, currents, and a frozen turbulence field with randomized coherent turbulent structures superimposed on the random, homogeneous background turbulence. All simulations are performed on the IEA-10MW \cite{osti_1529216} offshore reference wind turbine. TurbSim simulations for turbulence generation are performed with a grid resolution of $40$ points in a $1.16 \mathrm{D} \times 1.16 \mathrm{D}$ domain, $\mathrm{D}$ being the rotor diameter. The total simulation duration is \SI{900}{\second}, out of which the first \SI{300}{\second} is discarded to exclude the initial transient.

\subsection{Test Cases}
The input features, along with the variable bounds, are listed in \cref{table:aerohydro_variables}. The variable bounds for the aerodynamic inputs are referenced from Dimitrov et al. \cite{Dimitrov2018FromDatabases}. Power-law exponent and turbulence intensity are functions of wind speed. $R$ and $z$ are the rotor radius and the hub height, respectively. The wave scatter diagram for wave height and period is based on data provided by the SIMAR point 4038006, from the Spanish Ports Authority, as reported by Vigara et al. \cite{fernando_vigara_2020_4056779}. The first-order waves are modeled using the JONSWAP Spectrum in HydroDyn.  Sobol sequence is used where samples are uniformly distributed and pseudo-random sampling elsewhere. The random seeds are not included as training variables and are drawn using pseudo-random integer sampling in NumPy. Therefore, each sample is associated with a unique set of random seeds, and there are no repetitions. A total of $2491$ samples are generated.

    \begin{table}[!ht]
        \begin{center}
        \caption{\label{table:aerohydro_variables} \centering Variables and variable bounds for the aero-servo-hydro-elastic case. The feature vector is $\boldsymbol x = (u, TI, \alpha, H_\mathrm{s}, T_\mathrm{p}, W_\mathrm{dir})^\mathrm{T}$.}

        \begin{tabularx}{\linewidth}{LMN}
        \br
        Variable & Lower Bound & Upper Bound\\
        \mr
        Wind speed ($u$) [\si{\meter\per\second}] & $4$ & $25$ \\
        Turbulence intensity ($TI$) [\si{\percent}] & $2.5$ & $\frac{18}{u} (6.8+0.75u +3(\frac{10}{u})^2)$ \\
        Power law exponent ($\alpha$) [$-$] & $0.15 - 0.23(\frac{u_{max}}{u})(1-(0.4 \log \frac{R}{z})^2)$ & $0.22 +0.4(\frac{R}{z})(\frac{u_{max}}{u})$ \\
        Turbulence random seed [$-$] & $-50000$ & $50000$ \\
        Wave significant height ($H_\mathrm{s}$) [\si{\meter}] & $0$ & $6$ \\
        Wave time period ($T_\mathrm{p}$) [\si{\second}] & $1$ & $21$ \\
        Wave direction ($W_\mathrm{dir}$) [$-$] & $\SI{-180}{\degree}$ & $\SI{180}{\degree}$ \\
        Wave random seed 1 [$-$] & $-50000$ & $50000$ \\
        Wave random seed 2 [$-$] & $-50000$ & $50000$ \\
        \br
        \end{tabularx}
        \end{center}
    \end{table}
\subsection{Machine Learning Framework}
\begin{figure}[ht]
  \centering
  \includegraphics[page=1,width=0.8\textwidth]{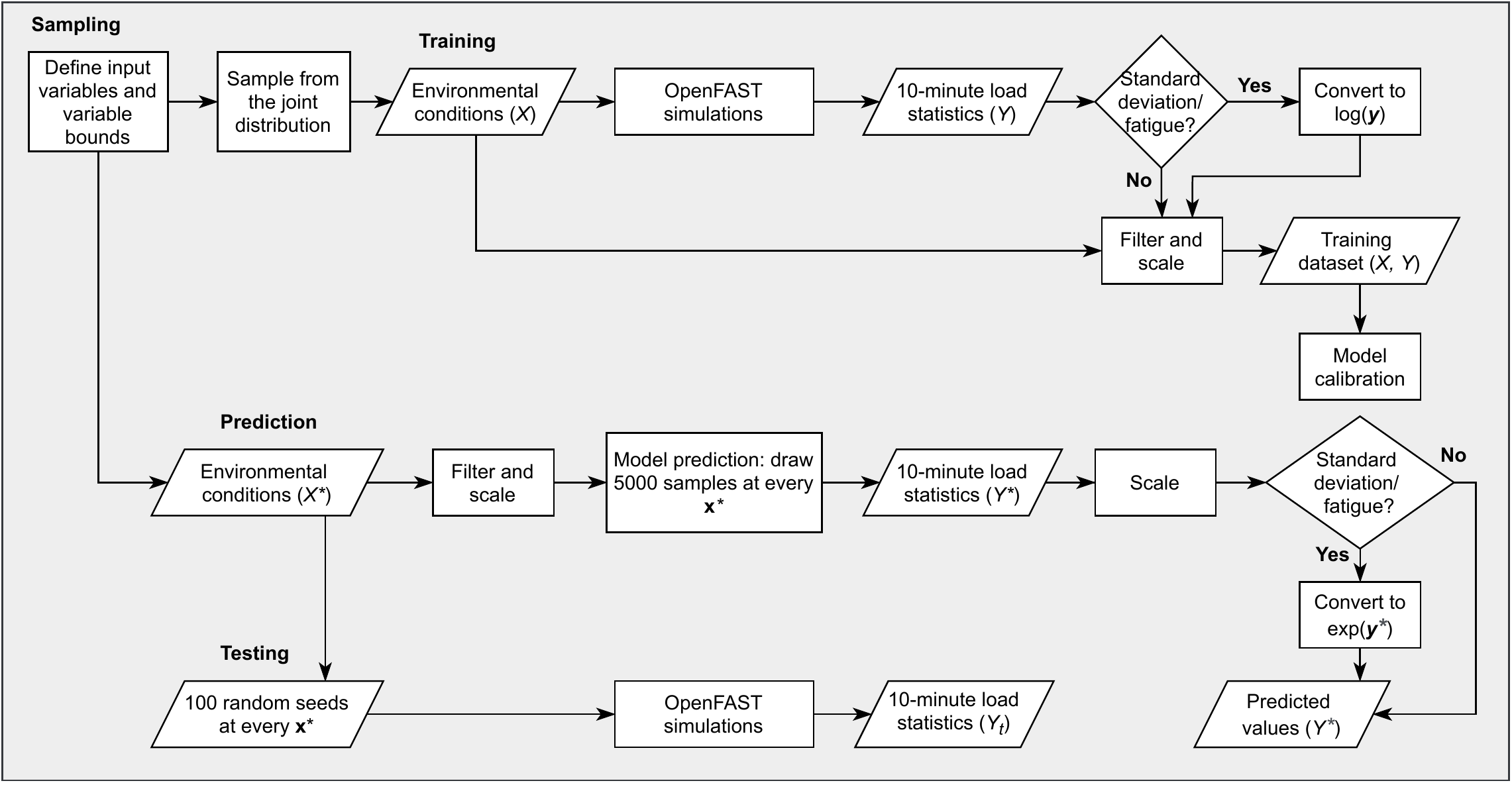}\hspace*{.01\textwidth}%
\caption{\label{fig:ml_framework}Schematic of the machine learning framework.}
\end{figure}

A schematic of the machine learning framework to train and test the model is shown in \cref{fig:ml_framework}. The surrogate model is trained on pairs of environmental conditions and $10$-minute load statistics that include the average, minimum, maximum, fatigue, and standard deviation of loads. We ensure that positive quantities like the standard deviation of the loads or the fatigue loads always return a positive prediction from the surrogate model by assuming them to be lognormally distributed. A Z-score filter with a threshold of $3\sigma$ is used for outlier detection. Outliers could result from a nonphysical combination of the inputs or discrete controller actions like turbine shut down at high wind speeds. The values of $\boldsymbol x$ and $y$ are scaled based on the standard deviation or the range.
 
We cannot compare point predictions from OpenFAST to point predictions from the surrogate because we would be comparing two random samples from the distributions. Instead, we compare the conditional distribution. To do that, a case-study is designed by repeating the OpenFAST simulations 100 times with the fixed inputs but unique random seeds at  $u = 6, 10, 14, 18, $ and $\SI{22}{\meter\per\second}$. It does not give a complete picture of the performance, but the validation of probabilistic approaches using a global accuracy metric is still an open question. 

\section{Results and Discussion}
\label{sec:results}
The methods introduced in \cref{sec:methodology} are calibrated on the cases presented in \cref{sec:setup}. The GPR model is trained on $2000$ sample points, and the H-GPR model is trained on $2000$ points with $1000$ inducing points. The computational cost of GPR was negligible, whereas the H-GPR model takes somewhere in the order of \SI{10}{\minute} on the same machine to train a single load channel. 

The pdf for H-GPR is generated by sampling $5000$ pairs of points $(f^{\star}, g^{\star})$ from each of the posterior latent functions $f(\boldsymbol x^{\star})$ and $g(\boldsymbol x^{\star})$, and using those to obtain a vector containing samples from $\mathcal{N}(\mu =f^{\star}, \sigma = e^{ g^{\star}})$. 
It, therefore, includes both data and model uncertainty. Similarly, the samples for GPR are drawn from the normal distribution with the modeled mean and marginal variance at $\boldsymbol x^{\star}$.

\begin{figure}[ht]
\centering
\captionsetup[subfigure]{justification=justified}
\begin{subfigure}{1.22\textwidth}
    \hspace*{-0.1\textwidth}
    \includegraphics[page=1,width=\textwidth]{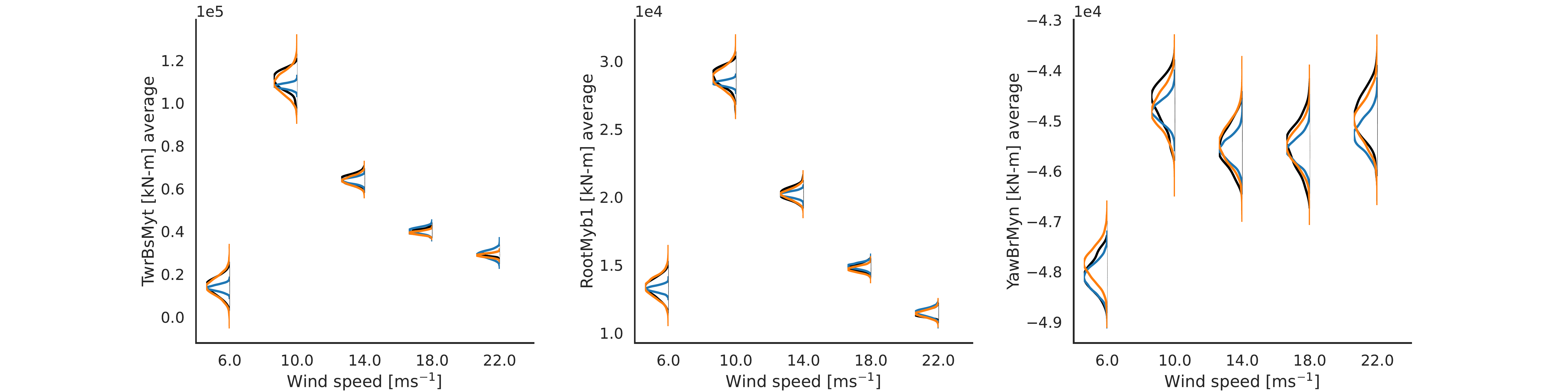}\hspace*{.01\textwidth}%
\caption{Predicted conditional distributions of load averages}
\label{fig:violinplot_average}
\end{subfigure}
\quad
\begin{subfigure}{1.22\textwidth}
    \hspace*{-0.1\textwidth}
    \includegraphics[width=\textwidth]{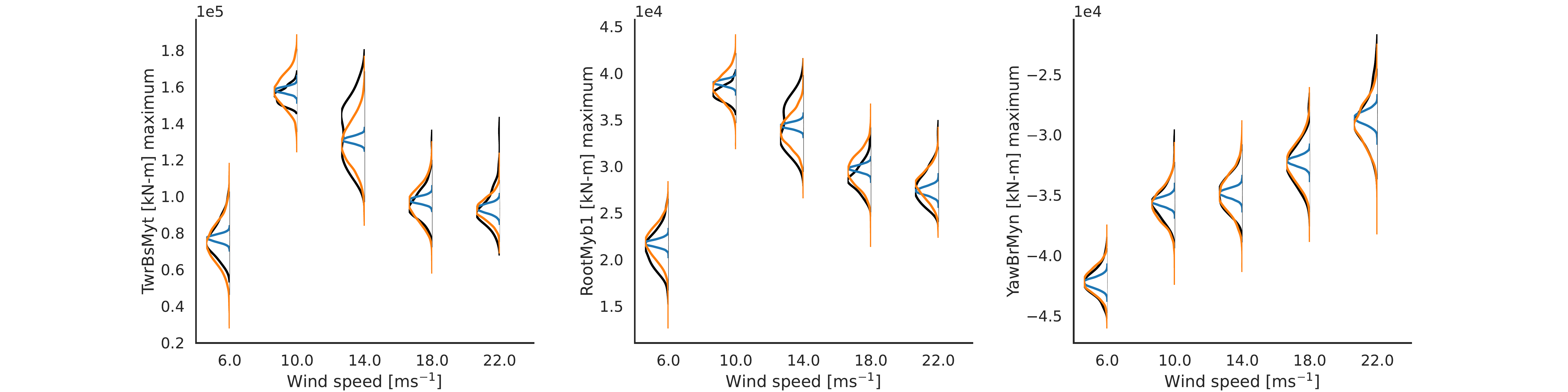}
\caption{Predicted conditional distributions of maximum loads}
\label{fig:violinplot_maximum} 
\end{subfigure}
\quad
\begin{subfigure}{1.22\textwidth}
    \hspace*{-0.1\textwidth}
    \includegraphics[width=\textwidth]{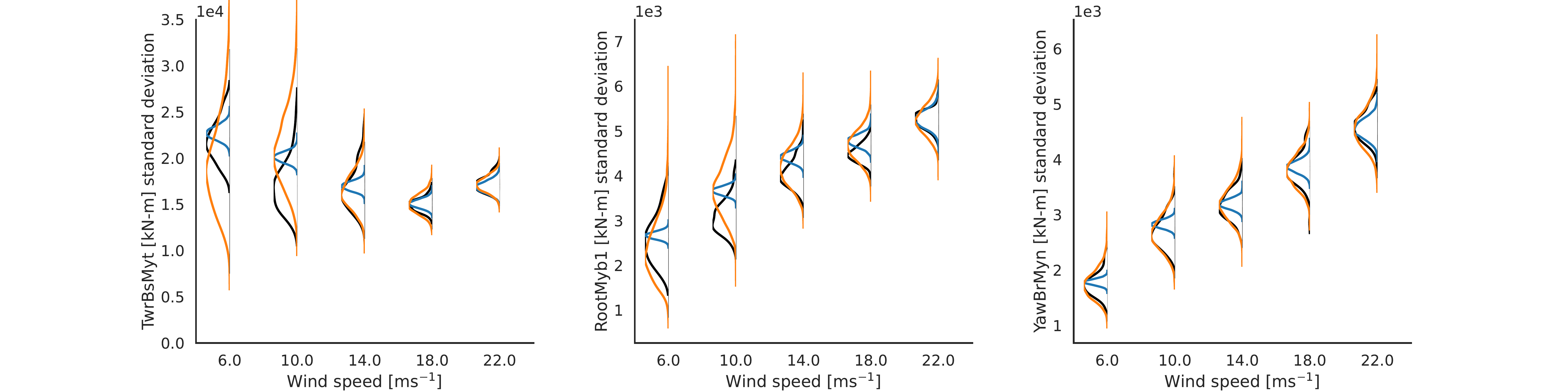}
\caption{Predicted conditional distribution of load standard deviation}
\label{fig:violinplot_stddev} 
\end{subfigure}
\caption{Comparison of the conditional pdf predicted by H-GPR (orange), GPR (blue) with OpenFAST reference (black) at $\boldsymbol x = (u, 12(0.75*u + 5.6)/u, 0.08, 1, 7, 0)^{\mathrm T}$. The tower base fore-aft moment \emph{TwrBsMyt} (left), blade root flapwise moment \emph{RootMyb1} (middle) and tower top fore-aft moment \emph{YawBrMyn} (right) are plotted. The reference pdf is calculated from 100 replications of the full-order model.}
\label{fig:violinplot}
\end{figure}

\begin{figure}[ht]
\centering
\begin{subfigure}{1\textwidth}
    \includegraphics[page=1,width=\textwidth]{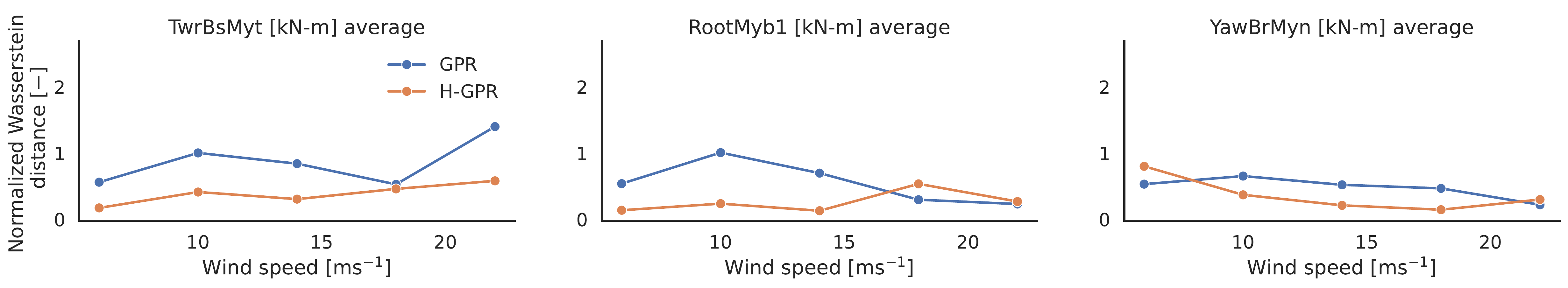}\hspace*{.01\textwidth}%
\caption{Load averages}
\label{fig:ws_average}
\end{subfigure}
\quad
\begin{subfigure}{1\textwidth}
    \includegraphics[width=\textwidth]{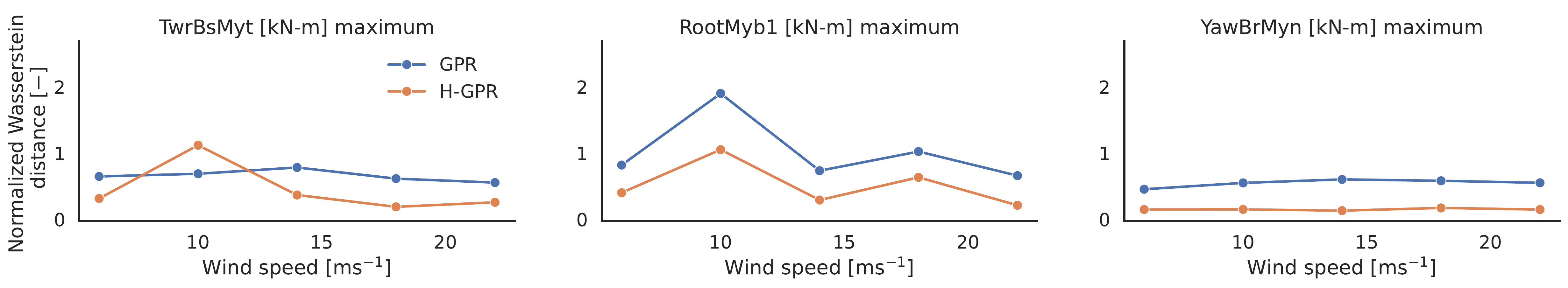}
\caption{Maximum loads}
\label{fig:ws_maximum} 
\end{subfigure}
\quad
\begin{subfigure}{1\textwidth}
    \includegraphics[width=\textwidth]{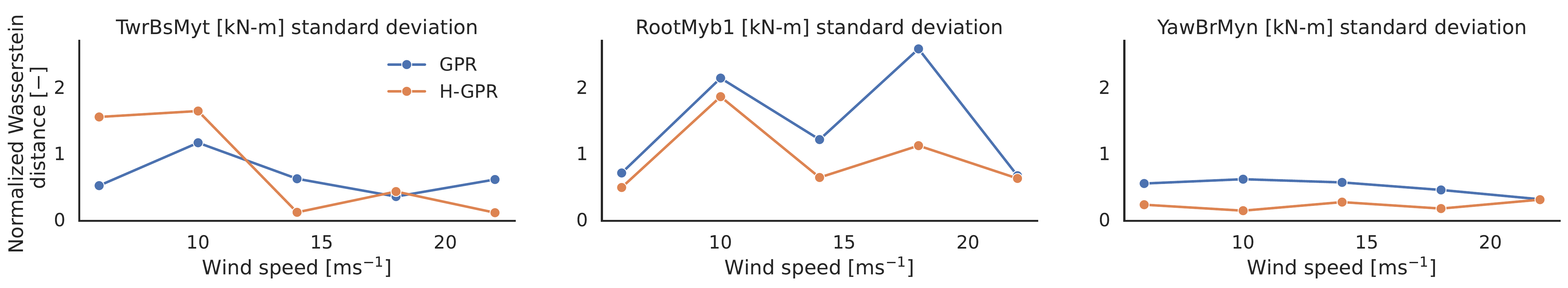}
\caption{Load standard deviation}
\label{fig:ws_stddev} 
\end{subfigure}
\caption{Normalized 1-Wasserstein distance between the OpenFAST reference and H-GPR (orange), as well as the reference and GPR (blue) at $\boldsymbol x = (u, 12(0.75*u + 5.6)/u, 0.08, 1, 7, 0)^{\mathrm T}$. The distances are plotted for the tower base fore-aft moment \emph{TwrBsMyt} (left), blade root flapwise moment \emph{RootMyb1} (middle) and tower top fore-aft moment \emph{YawBrMyn} (right).}
\label{fig:ws}
\end{figure}

In the paper we focus on the average, maximum and standard deviation (used as a proxy for fatigue loads) of the loads at the tower bottom (TwrBsMxt), blade root-flapwise (RootMyb1) and tower top (YawBrMyn). \Cref{fig:violinplot} shows the predicted pdf at $\boldsymbol x = (u, 12(0.75*u + 5.6)/u, 0.08, 1, 7, 0)^{\mathrm T}$. The difference between the conditional probability distributions is quantified using the normalized 1-Wasserstein distance \cite{panaretosWasserstein2019}. 
For two univariate distributions $P$ and $Q$, with distribution functions $F$ and $G$, the  1-Wasserstein distance is defined as,
\begin{equation*}
    W_1(P,Q) = {\int_{0}}^1 |F^{-1}(u)-G^{-1}(u)| du
\end{equation*}
and normalized by the standard deviation of the reference distribution,
\begin{equation*}
    d_{W_1}(P,Q)=\frac{W_1(P,Q)}{\sigma(P)}
\end{equation*}
Where, $F^{-1}$ and $G^{-1}$ are the quantile functions respectively. 
\Cref{fig:ws} shows the values of $d_{W_1}$ corresponding to the distribution functions in \cref{fig:violinplot}.
A qualitative comparison of the emulated distributions and OpenFAST in \cref{fig:violinplot_average} shows excellent predictions by H-GPR. The difference in the variance across wind speeds is not significant as averaged loads are less susceptible to stochastic inflow, yet it is clear that the heteroscedasticity is well approximated in all three load channels. The predicted mean also, therefore, shows better agreement in the case of H-GPR, especially around \SI{10}{\meter\per\second}. The maximum tower base fore-aft moment and blade root flapwise moment, \cref{fig:violinplot_maximum} exhibit strong gradients in the variance and mean around \SI{14}{\meter\per\second}. H-GPR is able to capture the trend correctly; however, the variances predicted by the model do not seem to change as rapidly as the reference. $d_{W_1}$ values indicate a good fit at mid-high wind speeds for all load channels, where the maximum loads are most critical from an engineering perspective. The maximum tower top fore-aft moment distributions are accurately predicted by H-GPR across all wind speeds, with $d_{W_1}  \approx 0.15$.

The load standard deviations at low wind speeds at the tower bottom and blade root show the largest inconsistency between H-GPR predictions and the reference. There are many reasons this behavior can be expected. First, H-GPR works with a few assumptions, such as modeling a Gaussian likelihood and using a Gaussian distribution to parametrize the approximated posterior. These assumptions are rigid and may not be valid throughout the domain. Any skewness in the conditional distribution, for instance, would result in over-prediction of the variance of the normal or lognormal distribution where H-GPR is expected to fit the data. Second, the model is trained on a relatively small training dataset, for such a high-dimensional case, as the computational cost of training the model scales rapidly, despite the sparse implementation. Insufficient data could result in an under-fitted variance function that cannot capture the local gradients adequately. Third, the reference distribution is generated with 100 points that may not result in a converged reference pdf. Finally, we are only able to visualize the variation of the pdf with respect to $u$, while the loads are influenced by the other features as well.

\section{Conclusion}
This paper presented a Bayesian approach to modeling the probability distribution function of offshore wind turbine loads. On account of the heteroscedasticity observed in the 10-minute load statistics calculated using a stochastic simulator, the sparse heteroscedastic Gaussian process regression's performance, modeled within the chained Gaussian process framework, is assessed for a specific test case. The results are compared against the standard Gaussian process regression for reference. Overall, within the limits of the number of training points and the assumptions in modeling, H-GPR can reproduce the conditional distributions at various inflow conditions with low error, quantified by the normalized Wasserstein distance. While the expected value of the distribution is not affected by the heteroscedastic model, the prediction of the variance shows a significant improvement over GPR. The method suffers from the curse of dimensionality and is unable to model very high gradients in the variance, especially those driven by hydrodynamic loads (at low wind speeds) or controller actions. 
However, the excellent predictions at mid to high wind speeds provide the user with a useful tool should they be interested in uncertainty quantification of site-specific fatigue or extreme loads in offshore wind turbines. From an industrial point of view, future research includes determining the value of probabilistic surrogates in estimating extreme events using the statistical information from the predictions rather than using very conservative design safety factors. The extension of the model to a high-dimensional problem such as floating offshore wind turbines is also of particular interest. 

\ack
The project has received funding from the European Union’s Horizon 2020 research and innovation program under grant agreement No. 860737 (STEP4WIND project, step4wind.eu).


\section*{References}
\vspace{1mm}

\begin{thebibliography}{17}
\bibitem{IEC_fixed} IEC 61400-3-1 Wind energy generation systems - Part 3-1: Design requirements for fixed offshore wind turbines IEC Geneva, CH

\bibitem{Muller2018ApplicationTurbines} M\"uller K and Wen Cheng P 2018 \textit{Wind Energy Science} \textbf{3} 149–162

\bibitem{Zwick2015TheDaniel} Zwick D and Muskulus M 2015 \textit{Wind Energy} \textbf{18} 1421–1432
 
\bibitem{Zhu2020Replication-basedDistributionsb} Zhu X and Sudret B 2020 \textit{International Journal for Uncertainty Quantification} \textbf{10} 249–275

\bibitem{Abdallah2019ParametricTurbines} Abdallah I, Lataniotis C and Sudret B 2019 \textit{Probabilistic Engineering Mechanics} \textbf{55} 67–77

\bibitem{Teixeira2017AnalysisSurfaces} Teixeira R, O’Connor A, Nogal M, Krishnan N and Nichols J 2017 \textit{Procedia Structural Integrity} \textbf{5} 951–958

\bibitem{Gasparis2020SurrogateFarm} Gasparis G, Lio W H and Meng F 2020 \textit{Energies} \textbf{13} 6360

\bibitem{Dimitrov2018FromDatabases} Dimitrov N, Kelly M C, Vignaroli A and Berg J 2018 \textit{Wind Energy Science} \textbf{3} 767–790

\bibitem{Saul2016ChainedProcesses} Saul A D, Hensman J, Vehtari A and Lawrence N D 2016 Chained Gaussian Processes \textit{Proc. of the 19th Int. Conf. on Artificial Intelligence and Statistics (Proc. of Machine Learning Research vol 51)} ed Gretton A and Robert C C (Cadiz: PMLR) pp 1431–1440

\bibitem{books/lib/RasmussenW06} Rasmussen C E and Williams C K I 2006 \textit{Gaussian processes for machine learning} Adaptive computation and machine learning (Cambridge, Mass.: MIT Press)

\bibitem{Blei2017VariationalStatisticians} Blei D M, Kucukelbir A and McAuliffe J D 2017 \textit{Journal of the American Statistical Association} \textbf{112} 859–877

\bibitem{GPflow2017} Matthews A G d G, van der Wilk M, Nickson T, Fujii K, Boukouvalas A, León-Villagrá P,
Ghahramani Z and Hensman J 2017 \textit{Journal of Machine Learning Research} \textbf{18} 1–6

\bibitem{openfast_web} https://github.com/openfast accessed: 2022-01-20

\bibitem{openfast_jonkman} Jonkman J 2013 The new modularization framework for the FAST wind turbine CAE tool
\textit{51st AIAA Aerospace Sciences Meeting including the New Horizons Forum and Aerospace Exposition} (Grapevine: AIAA)

\bibitem{osti_1529216} Bortolotti P, Tarres H C, Dykes K L, Merz K, Sethuraman L, Verelst D and Zahle F 2019
IEA Wind TCP Task 37: Systems Engineering in Wind Energy - WP2.1 Reference Wind Turbines Tech. Rep. NREL/TP-5000-73492 National Renewable Energy Lab. Colorado

\bibitem{fernando_vigara_2020_4056779} Vigara F, Cerdán L, Durán R, Muñoz S, Lynch M, Doole S, Molins C, Trubat P and Gunache R 2020 Design basis (1.0) Tech. rep.

\bibitem{panaretosWasserstein2019} Panaretos V M and Zemel Y 2019 \textit{Annual Review of Statistics and Its Application} \textbf{6} 405–431

\end{thebibliography}

\end{document}